\documentstyle[11pt,newpasp,twoside,epsf]{article}
\begin{document}
\title{{\tt SAURON} Observations of Disks in Spheroids}
\author{M.\ Bureau}
\affil{Columbia Astrophysics Laboratory, 550~W.\ 120th~St., 1027 Pupin
       Hall, Mail Code 5247, New York NY~10027, USA}
\author{R.\ Bacon, E.\ Emsellem}
\affil{CRAL, 9~Avenue Charles-Andr\'{e}, 69230 Saint-Genis-Laval, France}
\author{M.\ Cappellari, E.K.\ Verolme, P.T.\ de Zeeuw}
\affil{Sterrewacht Leiden, Niels Bohrweg~2, 2333~CA Leiden, Netherlands}
\author{Y.\ Copin}
\affil{Institut de Physique Nucleaire de Lyon, 4~rue Enrico Fermi,
       69622 Villeurbanne cedex, France}
\author{R.L.\ Davies, R.\ McDermid}
\affil{Physics Department, University of Durham, South Road, Durham
       DH1~3LE, UK} 
\author{H.\ Kuntschner}
\affil{ESO, Karl-Schwarzschild-Str.~2, 85748 Garching, Germany}
\author{B.W.\ Miller}
\affil{Gemini Observatory, Casilla~603, La Serena, Chile}
\author{R.F.\ Peletier}
\affil{Department of Physics and Astronomy, University of Nottingham,
       University Park, Nottingham NG7~2RD, UK}

%
%
\begin{abstract}
  The panoramic integral-field spectrograph {\tt SAURON} is currently
  being used to map the stellar kinematics, gaseous kinematics, and
  stellar populations of a large number of early-type galaxies and
  bulges. Here, we describe {\tt SAURON} observations of cold stellar
  disks embedded in spheroids (NGC~3384, NGC~4459, NGC~4526), we
  illustrate the kinematics and ionization state of large-scale
  gaseous disks (NGC~4278, NGC~7742), and we show preliminary
  comparisons of {\tt SAURON} data with barred galaxy N-body
  simulations (NGC~3623).
\end{abstract}

%
%
\section{Introduction}

\vspace*{-2.5mm}
The physical properties of early-type galaxies correlate with
luminosity and environment, as exemplified by the morphology-density
relation (Dressler 1980) and the dependence of color, metal content,
isophotal shape, nuclear cusp slope, and dynamical support on
luminosity (e.g.\ Bender \& Nieto 1990; Faber et al.\ 1997). It is
unclear to what extent these properties were acquired at the epoch of
galaxy formation or result from subsequent dynamical evolution.

Progress toward answering these and related issues requires a
systematic investigation of the kinematics and line-strengths of a
representative sample of early-type galaxies. The intrinsic shape,
internal orbital structure, and radial dependence of the mass-to-light
ratio are constrained by the stellar and gas kinematics (e.g.\ van der
Marel \& Franx 1993; Cretton, Rix, \& de Zeeuw 2000); the age and
metallicity of the stellar populations are constrained by the
absorption line-strengths (Gonzalez 1993; Davies, Sadler, \& Peletier
1993). {\tt SAURON} contributions to these two complementary
approaches are discussed in turn by Bacon et al.\ (2002) and Emsellem
et al.\ (2002) in these proceedings.

Here, we focus on the importance of disks in spheroids. In the current
context of hierarchical (i.e.\ merger-driven) structure formation, the
dichotomy in the properties of early-type galaxies (e.g.\ Faber et
al.\ 1997) is often explained by the mass ratio of the progenitor
galaxies (e.g.\ Barnes 1998; Naab, Burkert, \& Hernquist
1999). However, as revealed by {\tt SAURON}, the presence of cold
disks in a large number of spheroids suggests a non-negligible role
for dissipation, reminiscent of the old galaxy formation debate
between merging (Searle \& Zinn 1978) and dissipational collapse
(Eggen, Lynden-Bell, \& Sandage 1962).

After reviewing the core properties of the {\tt SAURON} instrument and
sample (\S~2), we present in \S~3 {\tt SAURON} observations of a few
cold central stellar disks embedded spheroids. We also illustrate the
emission line distributions and kinematics of large-scale gaseous
disks in \S~4, and compare generic N-body simulations of barred
galaxies with {\tt SAURON} data in \S~5.

%
%
\vspace*{-2.0mm}
\section{{\tt SAURON}: Instrument and Survey}

\vspace*{-2.5mm}
{\tt SAURON} ({\tt S}pectrographic {\tt A}real {\tt U}nit for {\tt
R}esearch on {\tt O}ptical {\tt N}ebulae) is a panoramic
integral-field spectrograph optimized for studies of the large-scale
kinematics and stellar populations of spheroids (Bacon et al.\ 2001,
hereafter Paper~I). In its low-resolution mode, it has a
$41\arcsec\times33\arcsec$ field-of-view sampled with
$0\farcs94\times0\farcs94$ lenslets, 100\% coverage, and high
throughput. The spectra cover 4810--5350~\AA\ with simultaneous sky
subtraction. Stellar kinematic information is derived from the Mg{\it
b} triplet and the Fe lines. The [OIII], H$\beta$, and [NI] emission
lines provide the morphology, kinematics, and limited information on
the ionization state of the ionized gas. The Mg{\it b}, H$\beta$, and
Fe5270 absorption lines constrain the age and metallicity of the
stellar populations.

The {\tt SAURON} survey targets a representative sample of 72 nearby
ellipticals, lenticulars, and early-type bulges constructed to be as
free of biases as possible while ensuring the existence of
complementary data (preferably space-based; $M_B \leq -18$). The
galaxies are further split into ``cluster'' and ``field'' objects and
populate the six $\epsilon-M_B$ planes uniformly (de Zeeuw et al.\
2002, hereafter Paper~II). By construction, the sample covers the full
range of environment, flattening, rotational support, nuclear cusp
slope, isophotal shape, etc.

%
%
\section{Stellar Disks Embedded in Spheroids}

We discuss below the stellar kinematics of 3 galaxies showing the
presence of a central stellar disk with varying strengths. The {\tt
SAURON} strategy is to map galaxies to one effective radius $R_e$,
which for nearly half the sample requires only one pointing. For the
largest galaxies, mosaics of two or three pointings reach 0.5~$R_e$
only. The mean stellar velocity $V$, the velocity dispersion $\sigma$,
and the Gauss-Hermite moments $h_3$ and $h_4$ (e.g.\ van der Marel \&
Franx 1993) are derived using the FCQ method (Bender 1990).

\subsection{NGC~3384}

NGC~3384 is a large SB0$^-$(s) galaxy in the Leo~I group ($M_B$=
--19.6). It forms a triple on the sky with NGC~3379 and NGC~3389 but
there is only marginal evidence for interactions. The light
distribution in the central $\approx20\arcsec$ is complex. The inner
isophotes are elongated along the major axis, suggesting an embedded
disk, but beyond $10\arcsec$ the elongation is along the minor-axis
(e.g.\ Busarello et al.\ 1996). The isophotes are off-centered at much
larger radii. NGC~3384 shows no emission lines, remains undetected in
HI, CO, radio continuum, and X-ray, but has IRAS detections at 12 and
100~$\mu$m (e.g.\ Roberts et al.\ 1991).

\begin{figure}
\plotone{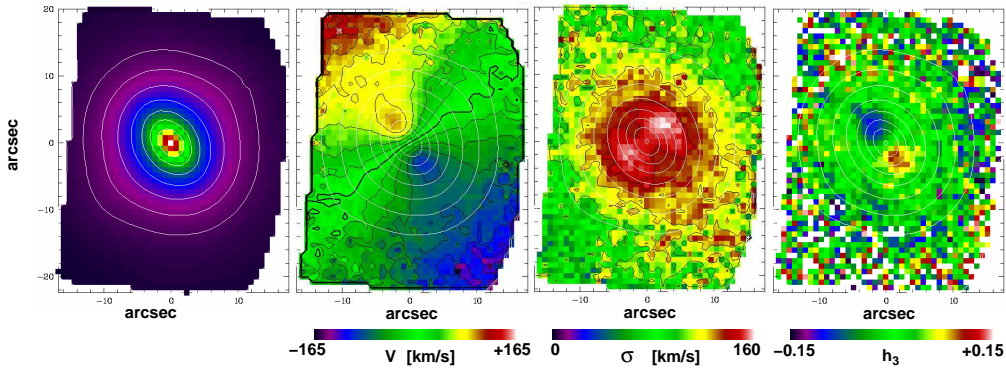}
\caption{{\tt SAURON} absorption-line measurements of the SB0 galaxy
 NGC~3384, based on a single pointing of $4\times 1800$~s. The
 effective spatial sampling is $0\farcs8\times0\farcs8$ and the seeing
 was $\approx2\farcs5$. a) Reconstructed total intensity $I$. b) Mean
 stellar velocity $V$. c) Stellar velocity dispersion $\sigma$. d)
 Gauss-Hermite moment $h_3$. The Gauss-Hermite moment $h_4$ displays
 little variation and is not shown. Isophotes are overlaid for
 reference.}
\end{figure}

Figure~1 displays the stellar kinematics of NGC~3384 and illustrates a
key advantage of {\tt SAURON}. Summing the flux in wavelength, the
galaxy surface brightness distribution is recovered and there is no
doubt about the relative location of the measurements. Figure~1 shows
that the bulge of NGC~3384 rotates regularly. Mean velocities increase
steeply along the major axis up to $r\approx4\arcsec$, then decrease
slightly, and rise again. No velocity gradient is observed along the
minor axis. The velocity dispersion map shows a symmetric dumb-bell
structure and the $h_3$ map is anti-correlated with $V$ also within
$r\approx4\arcsec$ (see Fisher 1997). All these facts point to the
presence of a cold, central stellar disk in NGC~3384.

\subsection{NGC~4526 and NGC~4459}

\vspace*{-1.0mm}
Many other galaxies in our sample show evidence of a central stellar
disk. NGC~3623 is discussed in Paper~II. Figure~2 shows two cases
where the disk appears to corotate with a central gaseous disk limited
by the dust lane (Rubin et al.\ 1997). NGC~4526 is a highly inclined
SAB0$^0$(s) galaxy in the Virgo cluster ($M_B$= --20.7). The stellar
disk is not visible in the reconstructed image but is evident in the
velocity and velocity dispersion fields. As in NGC~3384, the rotation
on the major axis first increases, then decreases, and increases again
in the outer parts, but the disk is much larger. It appears almost
edge-on, giving rise to an elongated depression across the (hot)
spheroid in the velocity dispersion map, and completely overwhelming
the central peak.

\begin{figure}
\plotone{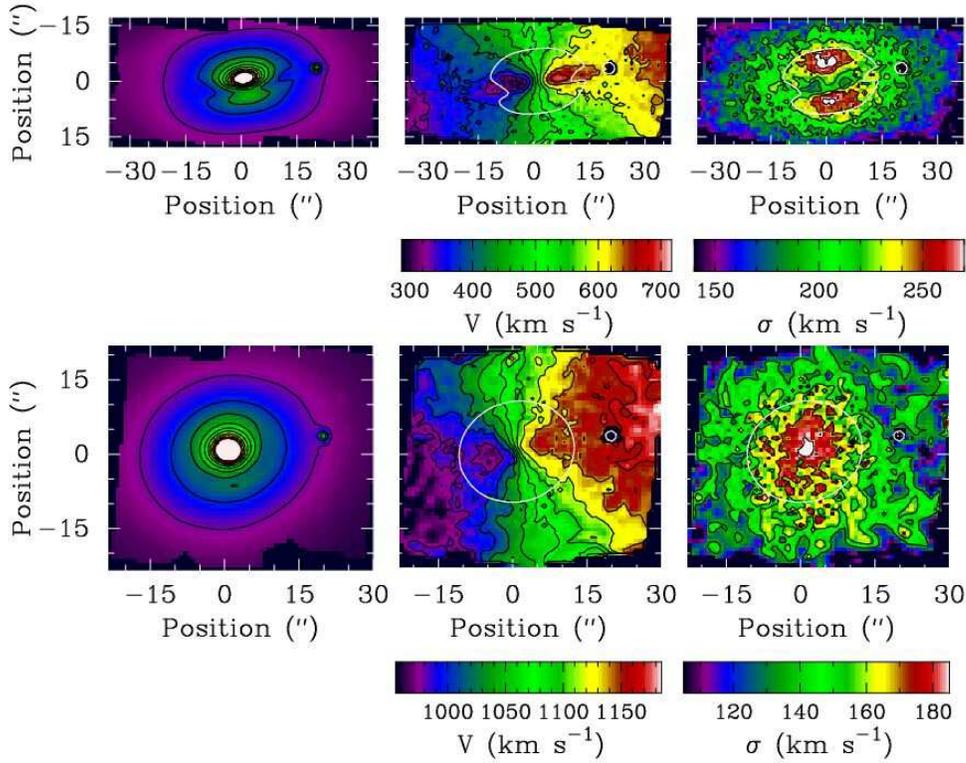}
\caption{{\tt SAURON} absorption-line measurements of the SB0 galaxies
NGC~4526 and NGC~4459, both based on two overlapping pointings. Top
row: Reconstructed total intensity, mean stellar velocity, and
velocity dispersion fields of NGC~4526. Bottom row: Same for
NGC~4459.}
\end{figure}

NGC~4459 is another S0$^+$(r) galaxy ($M_B$= --20.0) in Virgo. The
same behavior as in NGC~3384 and NGC~4526 is observed along the major
axis (see also Peterson 1978) but the minimum is shallower. The
isovelocity contours are also less skewed, indicating a more face-on
or intrinsically thicker disk (or both). This is supported by the
absence of a clear disk signature in the velocity dispersion map. A
statistical analysis of the entire {\tt SAURON} sample will reveal the
true fraction and necessary conditions for spheroids to harbor disks.

%
%
\section{Gaseous Kinematics and Ionization Mechanisms}

\vspace*{-2.5mm}
We now illustrate the scientific potential of the {\tt SAURON} gaseous
data. Paper~II describes how the H$\beta$, [NI], and [OIII] emission
lines are disentangled from the absorption lines by means of a
spectral library and optimal template fitting. Results on the
non-axisymmetric gaseous disks of NGC~3377 and NGC~5813 are presented in
Papers~I and II, respectively.

\subsection{NGC~7742}

\vspace*{-1.0mm}
NGC~7742 is a face-on Sb(r) spiral ($M_B$= --19.8) in a binary
system and is among the latest spirals included in our sample. De
Vaucouleurs \& Buta (1980) identified the inner stellar ring; Pogge \&
Eskridge (1993) later detected a corresponding small, bright ring of
HII regions with faint floculent spiral arms. NGC~7742 possesses a
large amount of HI, molecular gas, and dust (e.g.\ Roberts et al.\
1991) and is classified as a LINER/HII object.

\begin{figure}
\plotone{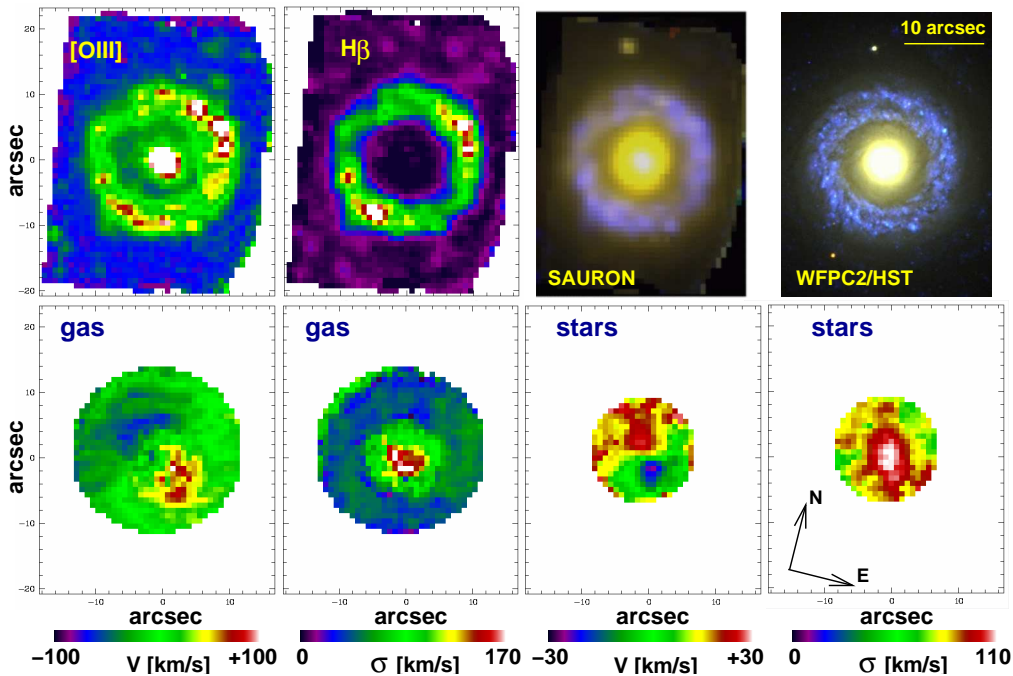}
\caption{{\tt SAURON} measurements of the stars and gas in NGC~7742
 from one pointing with seeing $\approx2\farcs0$. Top row:
 Emission-line intensity distributions of [OIII] and H$\beta$, followed
 by a reconstructed image composed of [OIII] and stellar continuum and
 a similar HST/WFPC2 image. Bottom row: Derived gas velocity and
 velocity dispersion fields, followed by the stellar velocity and
 velocity dispersion fields.}
\end{figure}

Figure~3 shows the [OIII] and H$\beta$ intensity maps, together with
the derived velocity and velocity dispersion fields. Most of the
emission is confined to a ring coinciding with the spiral arms.
H$\beta$ dominates in the ring (H$\beta$/[OIII]$\approx7-16$) but it
is much weaker in the center (H$\beta$/[OIII]$\approx1$). Also shown
in Figure~3 is a reconstructed image composed of [OIII] and stellar
continuum, and a similar image composed of HST/WFPC2 exposures. The
{\tt SAURON} data does not have HST's spatial resolution, but it does
show that our algorithms yield accurate emission-line maps. The main
surprise comes from the stellar and gas kinematics: the gas and stars
within the ring are counter-rotating.

\subsection{NGC~4278}

\vspace*{-1.0mm}
NGC~4278 is an E1-2 galaxy ($M_B$= --19.9) in the Virgo cluster. It
contains large-scale dust and a blue central point source (Carollo et
al.\ 1997). Its stellar rotation curve is peculiar, rising rapidly at
small distances from the nucleus and dropping to nearly zero beyond
$r\approx30\arcsec$ (Davies \& Birkinshaw 1988; van der Marel \& Franx
1993). NGC~4278 also contains a massive HI disk extending beyond
10~$R_e$. The HI velocity field is regular but has non-perpendicular
kinematic axes, indicating non-circular motions (Raimond et al.\ 1981;
Lees 1992).

\begin{figure}
\plotone{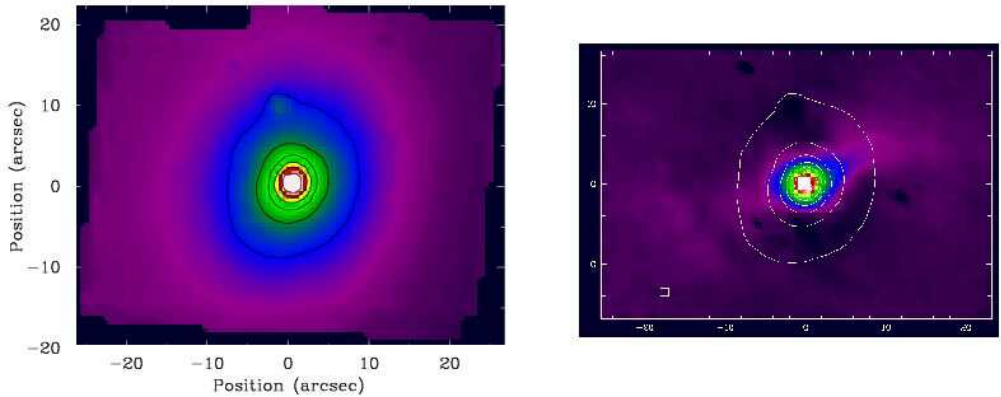}
\caption{{\tt SAURON} measurements of the stars and gas in NGC~4278,
 based on two pointings. Left: reconstructed stellar intensity. Right:
 [OIII] intensity. Isophotes are overlaid on both images. Scales are
 identical but fields-of-view differ slightly. The angular size of a
 {\tt SAURON} lenslet is indicated in the bottom left corner of the
 [OIII] image.}
\end{figure}

Figure~4 displays the reconstructed stellar intensity and [OIII] map
obtained with {\tt SAURON}. Despite the regular and well-aligned
stellar isophotes, the distribution of ionized gas is very extended
and strongly non-axisymmetric, similar to a bar terminated by ansae as
observed in spiral galaxies. The gaseous bar is reminiscent of the
dust structure observed at the center of Cen~A (Mirabel et al.\ 1999),
thought to have formed from the accretion of a gas-rich object, and
could be linked to the active nucleus (LINER/Sy1) and compact radio
core (Wrobel \& Heeschen 1984). Again, observations of the entire {\tt
SAURON} sample will reveal the fraction of spheroids with ionized gas
and its ionization and dynamical state.

%
%
\section{Comparing {\tt SAURON} and N-Body Data}

\vspace*{-2.0mm}
With full knowledge of phase-space, N-body simulations can easily be
projected and binned into three-dimensional data cubes. This allows an
easy comparison with stellar kinematic data from integral-field
spectrographs such as {\tt SAURON}, without the intermediate step of
deconvolution.

\subsection{NGC~3623}

In Figure~5, we compare {\tt SAURON} observations of NGC~3623, a
highly inclined SAB(rs)a galaxy in the Leo~I Group ($M_B$=--20.8;
Paper~II and references therein), with a generic barred galaxy N-body
simulation provided by E.\ Athanassoula (Athanassoula \& Misiriotis
2002; see also Bureau \& Athanassoula, in prep.). This is for
illustration purposes only, as no attempt was made to match either a
timestep or specific simulation with the observations. A viewing angle
was roughly estimated by eye. The line-of-sight velocity profiles
extracted from the simulations were fitted with the same algorithm as
the observations (the latter after deconvolution with FCQ), allowing a
direct comparison, but only major axis profiles were extracted due to
low S/N.

\begin{figure}
\plotone{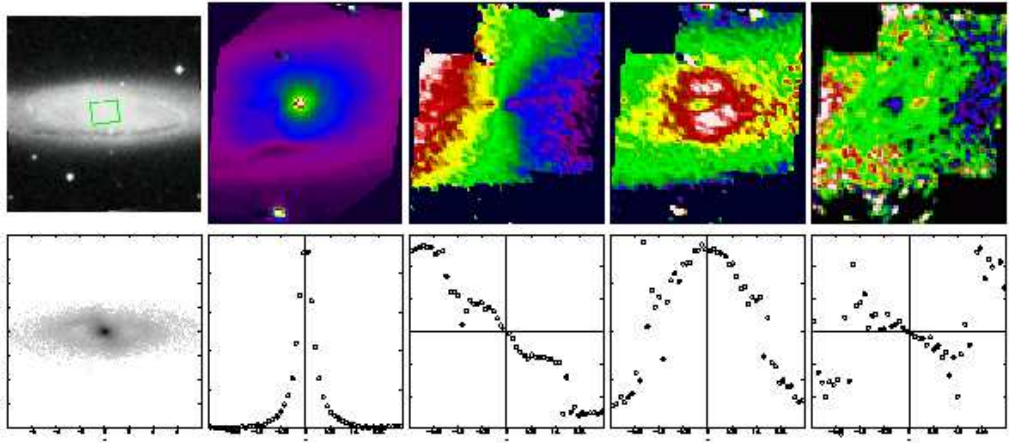}
\caption{Comparison of {\tt SAURON} observations with an N-body
 simulation. Top row: NGC~3623 large-scale image (Digitized Sky
 Survey), {\tt SAURON} reconstructed total intensity, mean stellar
 velocity, velocity dispersion, and $h_3$ fields, from two pointings. A
 single {\tt SAURON} footprint is overlaid on the DSS image (random
 orientation). Bottom row: Same for an N-body simulation, although only
 major axis profiles (cuts) are shown. The spatial axes of the N-body
 data were roughly scaled to NGC~3623; no attempt was made to scale the
 velocity data.}
\end{figure}

The rotation curve disk signature discussed above is clearly seen in
both datasets, but the simulation only shows a flattening of the
dispersion in the center, not a dip, and the $h_3$ profile is
correlated (rather than anti-correlated) with $V$ in the inner
parts. This suggests that the N-body central component is dynamically
hot and rather elongated, contrary to the central stellar disk of
NGC~3623. It remains to be seen if this is due to $x_2$ orbits. The
simulation also suggests that dissipation may indeed be necessary to
create the many stellar disks observed by {\tt SAURON} in spheroids.

%
%
\acknowledgments Support for this work was provided by NASA through
Hubble Fellowship grant HST-HF-01136.01 awarded by the Space Telescope
Science Institute, which is operated by the Association of
Universities for Research in Astronomy, Inc., for NASA, under contract
NAS~5-26555.

%
%

%
\end{document}